\documentclass[final,5p,times,twocolumn]{elsarticle}

\usepackage{amsmath,amssymb,amsfonts}

\usepackage{epsfig}
\usepackage{graphicx}
\usepackage{epstopdf}
\usepackage{amsmath,amssymb,amsfonts}
\usepackage{url}
\usepackage{lineno}
\usepackage[colorlinks=true,linktocpage=true,linkcolor=blue,citecolor=blue,allcolors=blue]{hyperref }
\usepackage{xspace}
\usepackage{multirow}
\usepackage{soul} 
\usepackage{float}
\usepackage{graphicx}
\usepackage{times}
\usepackage[normalem]{ulem}{
\usepackage{color}
\usepackage{cellspace}
\usepackage{array}
\usepackage{lipsum}

\journal{Physics Letter B}

\begin{document}

\begin{frontmatter}



\title{An Augmented QCD Phase Portrait: Mapping Quark-Hadron Deconfinement for Hot, 
Dense, Rotating Matter under Magnetic Field}
\author[ad1,ad2]{Gaurav Mukherjee}
\ead{phy.res.gaurav.m@gmail.com}
\author[ad1,ad2]{D. Dutta\corref{cor1}}
\ead{ddutta@barc.gov.in}
\cortext[cor1]{Corresponding author}
\author[ad1]{D. K. Mishra}
\ead{dkmishra@barc.gov.in}
\address[ad1]{Nuclear Physics Division, Bhabha Atomic Research Center, Mumbai 400085,India}
\address[ad2]{Homi Bhabha National Institute, Anushaktinagar, Mumbai - 400094, India}

\begin{abstract}
The quark-hadron transition that happens in ultra-relativistic heavy-ion collisions
is expected to be influenced by the effects of rotation and magnetic field, both present
due to the geometry of a generic non-head-on impact. We augment the conventional 
$T$--$\mu_B$ planar phase diagram for QCD matter by extending it to a 
multi-dimensional domain spanned by temperature $T$, baryon chemical potential 
$\mu_B$, external magnetic field $B$ and angular velocity $\omega$. Using two 
independent approaches, one from a rapid rise in entropy density and another dealing 
with a dip in the squared speed of sound, we identify deconfinement in the framework 
of a modified statistical hadronization model. We find that the deconfinement 
temperature $T_C(\mu_B,~\omega,~eB)$ decreases nearly monotonically with increasing 
$\mu_B,~\omega $ and $eB$ with the most prominent drop (by nearly $40$ to $50$ MeV) 
in $T_C$ occurring when all the three quasi-control (via collision energy and centrality) 
parameters are  simultaneously tuned to finite values that are typically achievable 
in present and upcoming heavy-ion colliders. 
\end{abstract}


\end{frontmatter}


\section{Introduction}
\label{sec:intro}
The beginning of this century marks the first experimental production and detection 
of quark-gluon plasma, an extreme phase of quantum chromodynamics (QCD), in  
ultra-relativistic heavy ion collisions (HIC) at the Relativistic Heavy Ion Collider 
(RHIC), BNL, and the Large Hadron Collider (LHC), CERN~\cite{Bzdak:2019pkr}. The 
same partonic matter is believed to have filled the early universe in its first 
microseconds after the Big Bang~\cite{Rischke:2003mt}. Upcoming facilities to 
further probe this exotic phase under varying conditions are under development, e.g., 
at Nuclotron-based Ion Collider fAcility (NICA), JINR, Dubna; the Facility for 
Antiproton and Ion Research (FAIR), GSI, Darmstadt; Heavy-Ion program at Japan Proton 
Accelerator Research Complex (JPARC-HI), Japan and   High Intensity heavy-ion 
Accelerator Facility (HIAF), China. 

Apart from very high temperatures and densities, the femto-scale system may also 
sustain significant rotation (parametrized by the angular velocity or vorticity 
$\omega \approx (9\pm1)\times10^{21}s^{-1} \sim 0.03$ fm$^{-1} \sim 0.006$ GeV as 
experimentally discovered by the RHIC collaboration~\cite{STAR:2017ckg}, higher 
theoretical estimates yield $\omega \sim 0.1$ fm$^{-1} \sim 0.02$ 
GeV~\cite{Becattini:2015ska,Jiang:2016woz,Deng:2016gyh}) and a strong magnetic field 
background ($B \sim 10^{19}$ Gauss corresponding to $eB \sim 6 m^{2}_{\pi} \sim 0.12 
$ GeV$^2$ where $e$ is the elementary charge and $m_\pi$ is the pion mass
~\cite{Kharzeev:2007jp,Deng:2012prc}) within the fireball droplet produced in a 
typical off-central 
collision~\cite{Huang:2016rpp,Becattini:2020ngo,Fukushima:2016vix}. The extreme 
values of all these parameters (HIC fireballs are likely the hottest, densest, most 
vortical and embedded in the strongest magnetic fields, albeit briefly) are unique in 
the present-day universe. The only other conceivable physical situations that are 
qualitatively analogous might be the very early universe~\cite{PhysRevD.50.2421} and 
neutron star or magnetar cores~\cite{Alford:2007xm} and even some quasi-relativistic 
condensed matter systems~\cite{Miransky:2015ava}.

Thus a comprehensive treatment should include variables like angular velocity and 
magnetic field as additional parameters characterizing the hot and dense QCD matter. 
While certain effects of finite baryon chemical potential ($\mu_B$), angular velocity 
($\omega$) and magnetic field ($eB$), have been studied either separately or pairwise 
in the past, the interplay of all three and their combined impact on deconfinement 
and heavy-ion phenomenology seem to have been unexplored until 
now~\cite{Fukushima:2016vix,Liu:2017spl,Fujimoto:2021xix,Jiang:2016wvv}. The present 
work aims to advance in this direction by focusing on the rich phase structure of 
strongly interacting matter that manifests in the proposed multi-dimensional 
parameter space relevant for high-energy nucleus-nucleus collisions.
 
Here we adopt a modified hadron resonance gas (HRG) model, aka the statistical 
hadronization model~\cite{Andronic:2018nat,Garg:2013ata}. It is a unique tool for the 
kind of investigation of QCD matter that we need to make for the non-perturbative 
regime under study. The success of this model comes from accurately reproducing the 
particle abundances in HIC experiments~\cite{Andronic:2012ut}. Moreover, HRG results 
have been found to be in good agreement  with lattice QCD simulations for 
temperatures upto the deconfinement temperature 
$T_C$~\cite{Andronic:2012ut,Tawfik:2004sw}. It also  predicts the freeze-out points 
on the phase diagram that lie remarkably close to the deconfinement boundary for 
$\mu_B$ upto and a little beyond $0.25$ GeV where the so-called sign problem of 
lattice methods thwarts ab initio attempts~\cite{Fodor:2004nz}. The thermodynamic 
quantities computed in the model can be used to formulate criteria that we impose in 
order to locate, characterize and constrain the desired region where quark-hadron 
deconfinement occurs in the phase diagram, as will be elucidated below.

Deconfinement of hadronic matter when heated above a certain temperature range 
releases the quarks and gluons, the fundamental constituents of the theory. The 
resulting proliferation of the degrees of freedom in the system leads to a rapid 
increase in various thermodynamic quantities and this property when viewed in the 
Hagedorn picture gives us a means to estimate the deconfinement region in the phase 
diagram~\cite{Hagedorn:1965st,Fukushima:2010is}. The transition should also leave an 
imprint in the squared speed of sound \cite{Khaidukov:2018lor} in the form of a 
prominent dip at the phase transition as may be intuitively expected given that it 
is essentially a ratio of the entropy density to the heat capacity and the latter 
diverges at a first-order phase transition. 

In this Letter, we shall confine our investigation to the deconfinement (crossover) 
zone in the fully augmented QCD phase diagram for a parameter space that is most 
relevant for present and upcoming colliders (LHC, RHIC, NICA, FAIR). To accomplish 
this we are going to adopt two independent methods of estimating the deconfinement 
temperature both under the framework of the HRG model. Section~\ref{sec:landau_Q_SOL} 
discusses the physical picture describing our system subjected to constraints and 
relations emerging from Landau quantization and relativistic causality. In 
Sec.~\ref{sec:model}, we begin with a formulation of the extended HRG model with 
parallel rotation and magnetic field followed by a computation of the required 
thermodynamic quantities. In Sec.~\ref{sec:strategy}, we discuss our strategy to 
study those characteristics that serve as excellent proxies for the onset of 
deconfinement and we selectively work with physical quantities that do not have a 
contribution from quantum zero-point (vacuum) fluctuations. Imposing a working 
criterion to extract the temperature of deconfinement, the entropy density will yield 
an estimate of the transition temperature. We will also study the speed of sound and 
impose a distinct condition from which another estimate for the deconfinement 
temperature may be obtained. In Sec.~\ref{sec:results}, we corroborate these 
results and discuss several  implications. Finally,  we summarize our findings and 
outline directions that may be studied further in the future in Sec.~\ref{sec:sum}.  

\section{Landau quantization and causality bound under rotation}
\label{sec:landau_Q_SOL}

We consider a relativistic quantum gas rigidly rotating with angular velocity 
$\overrightarrow{\omega}$ that is embedded in a uniform magnetic field 
{$\overrightarrow{B}$ and  contained within a cylinder of radius $R$ so that 
$\overrightarrow{\omega}=(0, 0, \omega_z \equiv \omega)$ and {$\overrightarrow{B}=(0, 
0, B_z \equiv B)$} are parallel. In such a system composed of both charged as well as 
neutral particles the former couple with the background magnetic field while the 
latter do not. To have a clear physical picture and disentangle the different 
influences let us first discuss just the magnetic coupling of a particle of charge 
$Q$. The well-known Landau levels, $\varepsilon = \sqrt{p_z^2+p_{\bot}^2+m^2} = 
\sqrt{p_z^2+|QB|(2n+1-2s_z)+m^2}$, appear with degeneracy $N=\lfloor \frac{|QB|S}{2 
\pi} \rfloor$ where $S \sim R^2$ ($S = \pi R^{2}$, for our geometry) is the 
transverse area and $n=0,1,2,..$ . We should note here that 
$p_{\bot}=\sqrt{|QB|(2n+1-2s_z)}$ is valid for both Dirac fermions with $s_z=\pm 1/2$ 
\cite{Chen:2017xrj} and scalar bosons \cite{Liu:2017spl}. We shall assume the same 
formula to hold true for particles of higher spin as well. Landau quantization 
dominates as long as $N \gg 1$ or equivalently $\frac{1}{\sqrt{|QB|}} \ll R$, or in 
words, $\frac{1}{\sqrt{|QB|}} \equiv l_B$, the so-called magnetic length which is the 
characteristic length scale for  the  cyclotron orbits, should be sufficiently 
smaller than the system size, $R=l_{\text{system}}$. If $l_{\text{system}} \lesssim 
l_B$, the cyclotron motion is disturbed by the boundary and the degenerate Landau 
quantized spectra no longer apply. Thus when the caveat for Landau levels, 
$(\frac{1}{\sqrt{|QB|}}=l_{B}) \ll R$ is violated (in the $B=0$ limit, say) the 
transverse momentum $p_{\bot}$ becomes of order $\frac{1}{R}$ (due to the edges) and 
not $\frac{1}{l_B}$ \cite{Chen:2017xrj}. We note that our work will employ Landau 
quantization and thus it strictly applies for sufficiently strong magnetic fields (or 
equivalently a large enough system size) relevant for HIC, unless otherwise 
mentioned. 

Now, the introduction of a rigid, global rotation necessitates us to confine the 
system within a cylindrical boundary since at a distance $r$ from the rotation axis, 
the azimuthal velocity, $v = r \omega$, should be below the vacuum speed of light, 
$c=1$. This causality bound then mandates the system size to be below a threshold 
$\frac{1}{\omega}$ which in turn provides an upper bound for the magnetic length 
$l_B$ which has to be smaller than the system size $R$ for Landau quantization to be 
valid~\cite{Chen:2015hfc}. This implies
\begin{equation}
\begin{aligned}
1/\sqrt{|QB|} \ll R \leqslant 1/\omega.
\label{ineq}
\end{aligned}
\end{equation}
 
The parameter space we shall study in this Letter falls within the bounds set by 
these inequalities. As a result, the $B=0$ substitution in our formulae for charged 
particles, for example the Landau quantized spectrum in $\varepsilon$ and other 
relations introduced below, is not valid and separate formulae will be used for the 
neutral particles. Alternative approaches in the literature (see 
Refs.~\cite{Liu:2017spl,Chen:2017xrj,PhysRevD.96.096014}) use a modified Landau 
spectrum in recognition of the distortion in the energy levels due to the boundary. 
This only becomes important when $l_{\text{system}} \lesssim l_B$ which corresponds 
to a very narrow sliver ($0 < eB \lesssim 0.0064$ GeV$^{2}$ for $l_{B} \sim 
l_{\text{system}} = 12.5 $~GeV$^{-1} \text{or~} 2.5 \text{~fm}$) in the phase space 
that we will investigate. Also, these distortions occur due to the so-called edge 
states and become essentially irrelevant in the deep interior, i.e., $r \ll R$, of 
the system where only the bulk states 
dominate~\cite{Chen:2017xrj,PhysRevD.96.096014}.     

The phase space sum, required for the thermodynamic potential gets modified for the 
case of magnetic field (since for a fixed $s_{z}$, $p_{\bot}^{2} = |QB|(2n+1-2s_z)$, 
$dp_x dp_y \to 2 \pi p_{\bot} dp_{\bot} = 2 \pi  |QB| dn$) to look 
like~\cite{Chen:2015hfc,Endrodi:2013cs} 
\begin{equation}
\begin{aligned}
\int \frac{dp_{x} dp_{y}}{(2 \pi)^2} \to \frac{|QB|}{2 \pi} \sum_{n=0}^{\infty}
\label{dim_red_mag}
\end{aligned}
\end{equation}
and for rotation accompanying the magnetic field the Landau degeneracy 
($N=\frac{|QB|S}{2 \pi}$ with $ S = \pi R^{2} $) is lifted by the canonical orbital 
angular momentum quantum number $l$ thus \cite{Chen:2015hfc}
\begin{equation}
\begin{aligned}
\int \frac{dp_{x} dp_{y}}{(2 \pi)^2} \to \frac{1}{\pi R^2} \sum_{n=0}^{\infty} 
\sum_{l=-n}^{N-n}
\label{dim_red_mag_rot}
\end{aligned}
\end{equation}
These relations show a dimensional reduction of the phase space that effectively 
transforms $ \int d^{3}p $ to just $ \int dp_{z} $ and takes the particle dynamics 
from $(3+1)$-dimensions $\to$ $(1+1)$-dimensions. The transverse phase plane $\int 2 
\pi p_{\bot} dp_{\bot}$ collapses due to the discretization from Landau quantization 
and leaves only the longitudinal degree of freedom to be integrated over.

Under the constraints embodied in Eq.~\ref{ineq}, we shall find that the complete 
dispersion relation for charged particles under $\overrightarrow{\omega} . 
\overrightarrow{B} > 0$ is given by Landau levels, $\varepsilon = 
\sqrt{p_z^2+|QB|(2n+1-2s_z)+m^2}$, with an accompanying effective chemical potential 
induced by the rotation that lifts the Landau degeneracy, thus yielding $\varepsilon 
\to \varepsilon - q \omega (l+s_{z})$ where $q=+1$ for the particle or positive 
charge state and $q=-1$ for the anti particle or negative charge state 
\cite{Liu:2017spl,Chen:2015hfc}.

In the rest of this section we briefly review the calculations, explicitly performed 
in Ref.~\cite{Chen:2015hfc} for fermions and Ref.~\cite{Liu:2017spl} for bosons, that 
support the formulae and arguments made above. Under the coordinate transformation 
suitable for a rigid global rotation, all local quantities can be expressed as 
functions of the co-rotating coordinates, $x^{\mu}$, in the non-inertial rotating 
frame of reference instead of $\tilde{x}^{\mu}$ in the rest (lab) frame.
The corresponding metric can be read as  
\begin{equation}
 g_{\mu \nu} = \eta_{a b} \dfrac{\partial \tilde{x}^a }{ \partial x^\mu } 
 \dfrac{\partial \tilde{x}^b }{ \partial x^\nu}=
\begin{pmatrix}
1-(x^2+y^2)\omega^2 & y\omega & -x\omega & 0\\
y\omega & -1 & 0 & 0\\
-x\omega & 0 & -1 & 0\\
0 & 0 & 0 & -1
\end{pmatrix},
\end{equation}
with the Minkowskian metric taken as $\eta = \text{diag}(1,-1,-1,-1)$. To deal with 
the fermions we introduce the vierbein, $\eta_{a b} = e^{\mu}_{a}  e^{\nu}_{b} g^{\mu 
\nu}$, and adopt   
\begin{equation}
\begin{aligned}
e^{t}_{0}=e^{x}_{1}=e^{y}_{2}=e^{z}_{3}=1,~~~~ e^{x}_{0}=y \omega,~~~~ e^{y}_{0}=-x \omega,
\end{aligned}
\end{equation}
taking all other components zero.

The Dirac equation in a curved background space-time and gauge field is given by
\begin{equation}
\begin{aligned}
&[i\gamma^{\mu} (D_{\mu} + \Gamma_{\mu}) - m] \psi = 0,
\label{diraceq}
\end{aligned}
\end{equation}
where the covariant derivative is $ D_{\mu} \equiv \partial_{\mu} + i Q A_{\mu}$, 
$Q$ being the charge of the Dirac fermion. The $\Gamma_{\mu}$ term is the usually 
defined affine connection associated with the rotating frame and containing Dirac 
spin matrices $\sigma^{ij} = \frac{i}{2} [\gamma^{i}, \gamma^{j}]$. Choosing the 
symmetric gauge $A_{\mu}=(0,By/2,-Bx/2,0)$, we can have the explicitly written Dirac 
equation under parallel $\overrightarrow{B}$ and  $\overrightarrow{\omega}$.
\begin{equation}
\begin{aligned}
&[i\gamma^{0} (\partial_{t} - x \omega \partial_{y} + y \omega \partial_{x} -i 
\omega \sigma^{12}) +i \gamma^{1} (\partial_{x} +i Q B y/2 )\\ &  +i 
\gamma^{2}(\partial_{y} - i Q B x/2) + i \gamma^{3} \partial_{z} - m] \psi = 0
\label{dirac_expl}
\end{aligned}
\end{equation}
We have already discussed the energy dispersion for spin $s$ particles in a magnetic 
field at $\omega=0$, given by
\begin{equation}
\begin{aligned}
E^{2} = p_z^2+|QB|(2n+1-2s_z)+m^2
\end{aligned}
\end{equation}
with $n=0,1,2,...$. We can separate out the contributions from rotation in 
Eq.~\ref{dirac_expl} which are made of terms $ -i (\omega x \partial_{y} - \omega y 
\partial_{x}) + \omega \sigma^{12} = \omega (\hat{L}_z+\hat{S}_z) $. If the 
eigenvalues of the $\hat{L}_z$ and $\hat{S}_z$ are denoted by $l$ and $s_z$ and from 
the recognition that $ E+\omega(l+s_z) $ is the energy eigenvalue in the inertial 
frame we arrive at 
\begin{equation}
\begin{aligned}
&[E+  \omega(l+s_z)]^{2} = p_z^2+|QB|(2n+1-2s_z)+m^2
\end{aligned}
\end{equation}
Thus the two branches of the energy eigenmodes $E=\pm|E|$ corresponding to the 
particle and antiparticle states yield $\sqrt{p_z^2+|QB|(2n+1-2s_z)+m^2} \mp 
\omega(l+s_z)$ respectively. This result applies to spinless particles as 
well~\cite{Liu:2017spl}. We now have all the ingredients (Eq.~\ref{dim_red_mag_rot} 
and the above dispersion relation) to estimate the thermodynamic potential or free 
energy density \cite{Chen:2015hfc,Ebihara:2016fwa,endrodilec18} required for the HRG 
model formulation.

A physical picture reveals how the charged particles behave  in a manner reminiscent 
of the classical Hall effect. The angular generalization ($\phi=0$ to $2\pi$) of the 
linear Hall effect situation may be the mechanism via which a charge gradient or 
imbalance grows between the radially inner and outer regions of the circulating gas 
leading to a quasi-static or nearr-equilibrium system that can be legitimately 
treated in a statistically equilibrated ensemble for $\omega \ll \sqrt{|QB|}$.

We emphasize that as we are only interested in the bulk interior of the cylinder,
the finite-size effects (due to edge states) are neglected in the calculations for
the charged particles. For the regime that we operate in (strong magnetic field), 
there is negligible dependence on the local radial distance upto $r \sim 0.8 R$ 
since the edge states are localized near the boundary only 
\cite{Chen:2017xrj,PhysRevD.96.096014}. Schematically,  a small magnetic length 
allows the system to accommodate `tight' cyclotron orbits in the bulk that are 
intact from boundary interference. The opposite limit of large magnetic length ($B 
\to 0$) lets the cyclotron orbits `loosen up' (i.e., become less curved) and when 
they become comparable to the system size the boundary conditions become important. 
This implies that for the $B=0$ case we need to include the finite size effects and 
thus expect $r$ dependence. We shall see this explicitly in the next section.

\section{Statistical hadronization model with rotation and magnetic field}
\label{sec:model}

The composite ideal gas system in the HRG model consists of the charged baryons and 
mesons which are affected by the magnetic field (via Landau quantization) as well 
as the neutral particles that are not, if we neglect their anomalous magnetic 
moments. The corresponding thermodynamic formulae governing the charged and neutral 
particles thus look different as shown below.    

The free energy density with the vacuum term suppressed for charged baryons and 
mesons, as discussed in the previous section, is expressed as
\begin{equation}
\begin{aligned}
 f_{i,c}^{b/m} = \mp \frac{T}{\pi R^2} \int \frac{dp_z}{2 \pi}
 \sum_{n=0}^{\infty}\sum_{l=-n}^{N-n} \sum_{s_z=-s_{i}}^{s_{i}}
 \ln(1 \pm e^{-(\varepsilon_{i,c}-q_i\omega (l+s_{z})- \mu_{i})/T}) ,
 \label{eq_f_charged}
\end{aligned}
\end{equation}
where the dispersion relation contains the Landau levels
\begin{equation}
\begin{aligned}
\varepsilon_{i,c} =& \sqrt{p_{z}^{2}+m_{i}^{2}+ | Q_{i} B | ( 2n - 2s_{z} +1) }~~.
\label{eq_e_charged}
\end{aligned}
\end{equation}
This obeys the constraint in Eq.~\ref{ineq} as we use undistorted Landau levels. The 
free energy density for the neutral particles \cite{Fujimoto:2021xix} is given by
\begin{equation}
\begin{aligned}
f_{i,n}^{b/m} = \mp \frac{T}{8\pi^2} \int_{(\Lambda_{l}^{\text{IR}})^{2}} {{dp}^{2}_{r} \int {dp_z}  
\sum_{l=-\infty}^{\infty}\sum_{\nu=l}^{l+2s_{i}} J^{2}_{\nu}(p_{r}r)}\\ \times \ln(1 \pm 
e^{-(\varepsilon_{i,n}-(l+s_{i})\omega -\mu_{i})/T}) , 
\label{eq_f_neutral}
\end{aligned}
\end{equation}
where the free part of the energy dispersion is given by 
\begin{equation}
\begin{aligned}
\varepsilon_{i,n}= \sqrt{p_{r}^{2}+p_{z}^{2}+m_{i}^{2}}~~.
\label{eq_e_neutral}
\end{aligned}
\end{equation}

Here $Q_i$, $q_{i}=Q_{i}/|Q_{i}|$, $s_i$ and $m_i$ are the charge, sign of charge, 
spin and mass of the $i^{th}$ hadron and the subscripts $c$ and $n$ refer to charged 
particles and neutral particles respectively. The upper (lower)  signs correspond to 
the  baryons (mesons) as denoted by the  superscript $b$ ($m$). The chemical 
potential $\mu_{i} = Q_{B,i} \mu_{B} + Q_{e,i} \mu_{e} + Q_{S,i} \mu_{S} $ reflects 
the baryonic, electric charge and strangeness components of the $i^{th}$ particle.  
We have set $\mu_{e}=0$ and $\mu_{S}=0$ here for simplicity and shall report the 
consequences of charge conservation and zero net-strangeness conditions in succeeding 
works. 

In Eq.~\ref{eq_f_charged}, the Landau degeneracy is lifted by rotation and the 
degenerate quantum number is the canonical orbital angular momentum $l$. The 
summation $ \sum_{l=-n}^{N-n} $ over $l$ runs upto $N-n$. Since $N$ is a function of 
the magnetic field, differentiation of the free energy density with respect to the 
magnetic field requires us to take this into account. Similarly, in 
Eq.~\ref{eq_f_neutral}, there is an $\omega$-dependent infrared cutoff 
($\Lambda^{\text{IR}}_l$) in the $p_r$ integration $\int_{\Lambda^{\text{IR}}_{l}} 
{dp}^{2}_{r}$, where $\Lambda^{\text{IR}}_{l}=\xi_{l,1}\omega$ and $\xi_{l,1}$ is the 
first zero of the Bessel function.

We now comment on the conspicuous presence of the transverse phase space and 
configuration space coordinates $p_{r}$ or $p_{\bot}$ and $r$ in 
Eq.~\ref{eq_f_neutral}. As explained in Sec.~\ref{sec:landau_Q_SOL}, both of these 
are expected for the neutral particles which are insensitive to the magnetic field. 
Their trajectories are thus strongly affected by the centrifugation effect due to 
rotation and the finite boundary. This is not the case for Eq.~\ref{eq_f_charged} 
when the magnetic field is large enough. More specifically, the dimensional reduction 
of phase space from Landau quantization 
(Eqs.~\ref{dim_red_mag},\ref{dim_red_mag_rot}) leaves only $\int dp_{z}$ in Eq.~ 
\ref{eq_f_charged} whereas for the $B=0$ case in Eq.~\ref{eq_f_neutral}, $\int 
dp_{r}^{2}$ is retained. 

If we had included boundary effects for $B \neq 0$ (as done in 
Ref.~\cite{Chen:2017xrj} for $\omega = 0$), we would have $\varepsilon = 
\sqrt{p_z^2+2|QB|\lambda_{l,k}+m^2} $ (instead of $\sqrt{p_z^2+2|QB|n+m^2}$ for spin 
up), where $ \lambda_{l,k} $ represents a modified Landau level index in a 
finite-size system and its explicit form depends on the boundary condition at $ r = R 
$. With an imposed boundary the Landau wavefunction is also deformed by the 
finite-size effect and contains the confluent hypergeometric function. These 
wavefunctions $\Phi_{l}$ and $\Phi_{l+1}$ in that case reduce to Bessel functions in 
the limit $B \to 0$, i.e., $\Phi_{l}(\lambda_{l,k},\frac{|QB|r^{2}}{2}) \to 
J_{l}(\sqrt{2|QB|\lambda_{l,k}} r) $ and 
$\Phi_{l+1}(\lambda_{l,k}-1,\frac{|QB|r^{2}}{2}) \to 
J_{l+1}(\sqrt{2|QB|\lambda_{l,k}} r) $. We identify the argument 
$\sqrt{2|QB|\lambda_{l,k}} r$ with $p_{\bot} r = p_{r} r$. This justifies the 
appearance of the Bessel functions in Eq. \ref{eq_f_neutral} where boundary effects 
are accounted for.

There is no counterpart in Eq.~\ref{eq_f_charged} of the Bessel functions in 
Eq.~\ref{eq_f_neutral} on account of the following reason. In Eq.~\ref{eq_f_neutral} 
we work in cylindrical coordinates since it is the natural formulation suited to the 
geometry of the system, and the Bessel function arises from the weight in the 
Bessel-Fourier expansion. However, in a magnetic field obeying Eq.~\ref{ineq} the 
dimensional reduction in Eq.~\ref{dim_red_mag_rot} encapsulates the transverse phase 
plane and thus $p_r$ disappears in Eq.~\ref{eq_f_charged} along with the 
$J_{\nu}^{2}(p_{r} r)$ part. A possible justification for this may be that the 
wave-function should be exponentially localized around the axis ($r=0$) for large 
$eB$, and the Bessel function part could be approximated by unity at $r=0$.    

We note that in our approximation Eq.~\ref{eq_f_neutral} is not recoverable from 
Eq.~\ref{eq_f_charged} under the substitution $B=0$ simply because $B$ is not a small 
perturbation (no weak field limit $B \to 0$ exists) in Eq.~\ref{eq_f_charged}, in 
light of the constraint in the first inequality in Eq.~\ref{ineq}. We remark that 
this kind of a disconnected $B=0$ and strong $B \neq 0$ treatment applied to the HRG 
model has already been successfully employed in the literature (see 
Refs.~\cite{Fukushima:2016vix,Endrodi:2013cs,Bhattacharyya_2016,PhysRevC.99.024902}
). 
 
 We have taken fixed values of $r = 3$ GeV$^{-1}$ and $R = 12.5$ GeV$^{-1}$
(or $R=2.5$ fm, $\sim$ system radius for peripheral collisions at 
freeze-out~\cite{STAR:2017prc}), so as to have definite, unambiguous results valid 
for our idealized system. Since our principal aim here is to explore the phase 
structure, we bypass a thorough study of the radial dependence and possible 
inhomogeneous equations of state~\cite{Chernodub:2020qah}.

 For the system we are dealing with, the free energy density reads  

$f=\epsilon - T s - \mu_B n_B - \omega j - B m_B=-p$,  

where pressure $p$, energy density $\epsilon$, entropy density $s$, 
baryon number density $n_B$, magnetization $m_B$ and total angular 
momentum $j$ are the relevant observables.  
All the observables here satisfy simple differential relations, 
 $s = -\frac{\partial f }{ \partial T}, ~ n_B = -\frac{\partial f }{ \partial \mu_B},  
~j = -\frac{\partial f }{ \partial \omega},  ~  m_B = -\frac{\partial f }{ \partial B}$.
It is required to use Leibniz rule to carry out 
differentiation of the free energy with respect to the angular velocity while 
computing the total angular momentum. 
The squared speed of sound is defined as \( c_s^2 = 
\frac{\partial p} { \partial \epsilon} |_{(\mu_B, \omega, eB)} 
 = [\frac{\partial p }{ \partial T } |_{(\mu_B, \omega, eB)}]/ [\frac{\partial \epsilon }{ \partial T } |_{(\mu_B, \omega, eB)}] \) 
 and here comes out to be
\begin{equation}
\begin{aligned}
c_s^2
 =\dfrac{s} { \Big( T \dfrac{\partial s }{ \partial T} 
 + B \dfrac{\partial m_B }{ \partial T} + \omega \dfrac{\partial j }{ \partial T} + 
\mu_B \dfrac{\partial n_B }{ \partial T} \Big)\Big|_{(\mu_B, \omega, eB)} }~~.\\
\end{aligned}
\end{equation}
All required thermodynamic quantities can be computed from the free energy density 
and plugged in to obtain the desired results.

All hadrons, upto a mass of $1.5$ GeV and excluding those having spin-$3/2$, listed 
in the particle data group list of particles contained in the package of 
THERMUS-V3.0~\cite{THERMUS:2009} have been included in our HRG model treatment. The 
ultraviolet mass cut-off is taken to reduce numerical cost and the exclusion of the 
spin-$3/2$ sector has been implemented due an instability in its theory 
~\cite{Endrodi:2013cs,Rarita:1941mf,Velo:1969bt,Johnson:1960vt}. 

\section{Strategy to identify deconfinement}
\label{sec:strategy}

The first criterion that we employ to characterize the deconfinement transition and 
demarcate its location on the multi-dimensional QCD phase diagram is based on an 
argument regarding the physical reinterpretation of the Hagedorn limiting temperature 
concept~\cite{Hagedorn:1965st,Rafelski:2016hnq}. Historically, the original 
interpretation of the Hagedorn temperature as a limiting temperature within the 
hadronic bootstrap model framework was revised ~\cite{Cabibbo:1975ig} when it was 
realized that the same should be construed as the transition temperature to more 
fundamental degrees of freedom, namely the quarks and gluons. A drastic rise in 
thermodynamic quantities like the entropy density is expected during the 
deconfinement transition. Since this transition is of crossover type, we do not see a 
strict discontinuity or divergence but rather a sharp change within a narrow 
temperature window. We choose a working condition for deconfinement along these 
lines~\cite{Fujimoto:2021xix,Fukushima:2010is}. From an experimental standpoint too 
there is a strong indication that the hadro-chemical freeze-out curve determined by 
the so-called universal freeze-out conditions can act as a close proxy for the 
deconfinement band region in the phase diagram, and this is most accurate in the 
small to medium baryo-chemical potential range~\cite{Fodor:2004nz}. 

Another pragmatic way to deduce the onset of deconfinement involves the speed of 
sound which  has been studied to distinguish QCD phases in highly varied 
contexts~\cite{Annala:2019puf,He:2022kbc}. In contrast to the somewhat arbitrary 
condition imposed on the scaled entropy density to reach the $T_C$ estimate, the dip 
in the squared speed of sound, $c_s^2$ provides precise values for $T_C$ 
independently \cite{Borsanyi:2010cj}. If the phase transition  were strictly 
first-order then we would have expected a discontinuity at $T=T_C$ where $c_s^2$ 
should vanish~\cite{Khaidukov:2018lor}. The crossover nature of the quark-hadron 
transition leads to the relatively smooth and shallow minima of $c_s^2$ occurring at 
and near the deconfinement as seen in Fig.~\ref{fig:sos}. We shall now extend the 
evaluation of these observables to new regimes in parameter space using the 
generalized HRG model as detailed above. Extrapolating the validity of the conditions 
for deconfinement, we will be able to map out the QCD phases in a multi-dimensional 
domain.

\section{Results and discussion}
\label{sec:results}
We compute the entropy density numerically and impose the criteria $ s/T^3 =$ $4$ -- 
$7$  to constrain the range within which the deconfinement transition occurs most 
dramatically (rapid rise in thermodynamic quantities). This leads to the 
deconfinement bands shown in Fig.~\ref{fig:bands} for the scenarios without ($eB$ = 
0) and with ($eB=0.15$ GeV$^2$) magnetic field. We have taken $\mu_e=0$ and $\mu_S=0$ 
for simplicity.  
\begin{figure}[h]
\includegraphics[width=70mm]{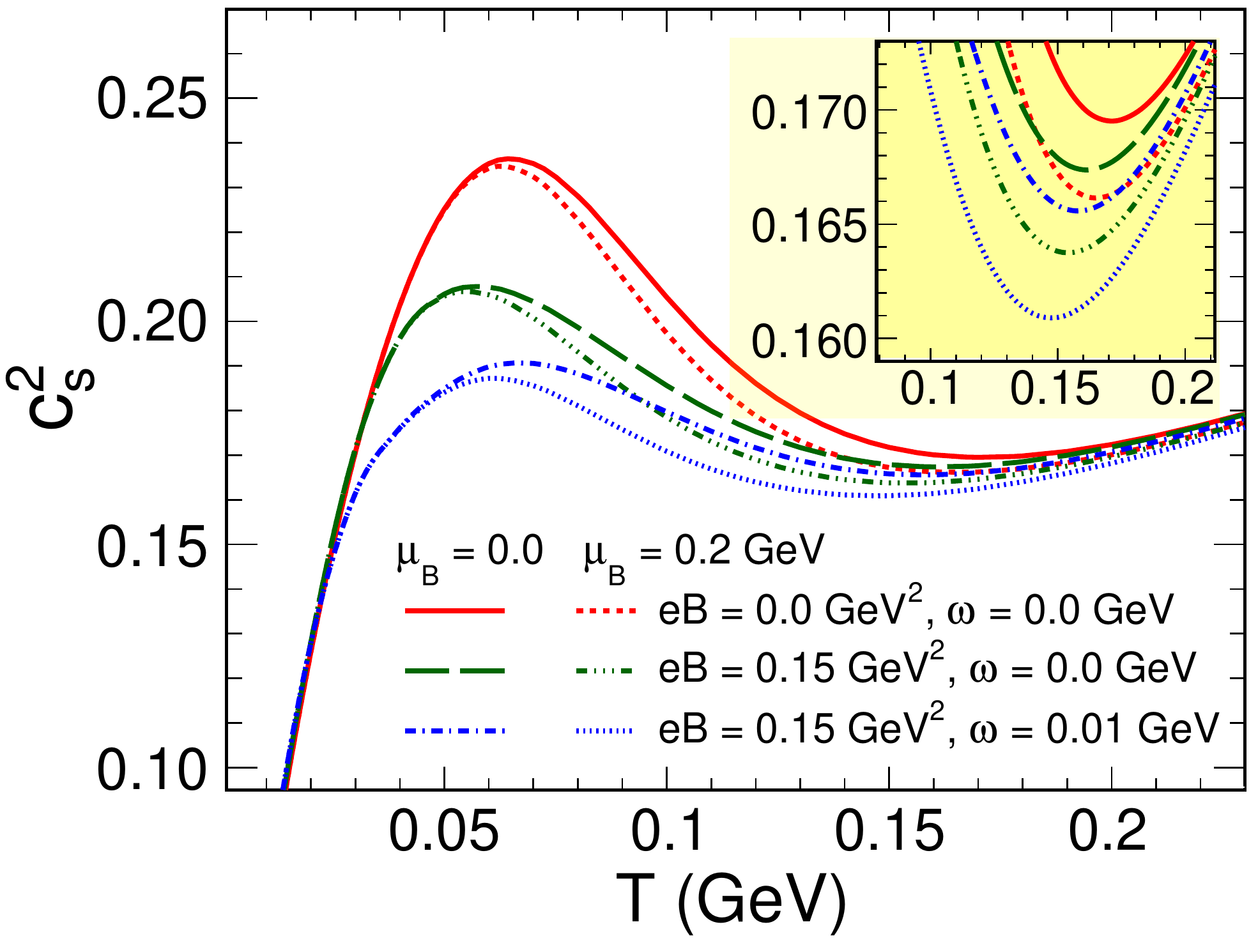}
\caption{\label{fig:sos}The variation of the squared speed of sound as a function of 
temperature with a magnified view of the region where the minima occur in the inset.}
\end{figure}
\begin{figure}[ht]
\includegraphics[width=0.5\textwidth]{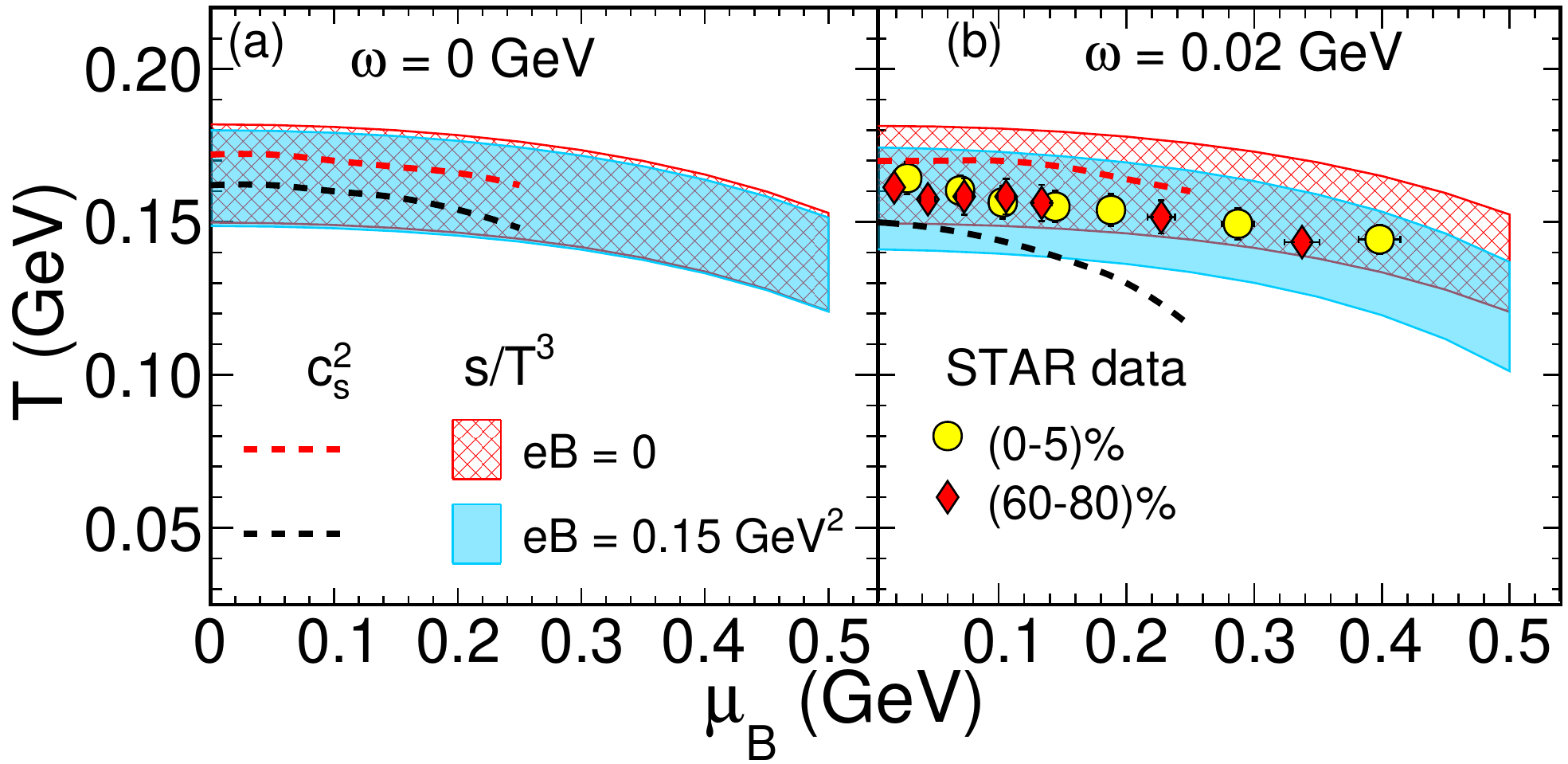}
\caption{\label{fig:bands} QCD phase diagrams, $T$ vs. $\mu_B$ for $eB=0$ 
(red band or  curve) and $eB=0.15$ GeV$^2$ (blue band or  curve)
and  (a) for $\omega=0$ GeV and (b) for $\omega=0.02$ GeV. The deconfinement 
transition zones depicted as (i) bands constrained by $s/T^3 = 4$ (lower edge) 
and $7$ (upper edge) , and (ii) curves obtained from the minima of $c_s^2$ 
vs. $T$ as shown in Fig.~\ref{fig:sos}.}
\end{figure}
\begin{figure}[h]
\includegraphics[width=70mm]{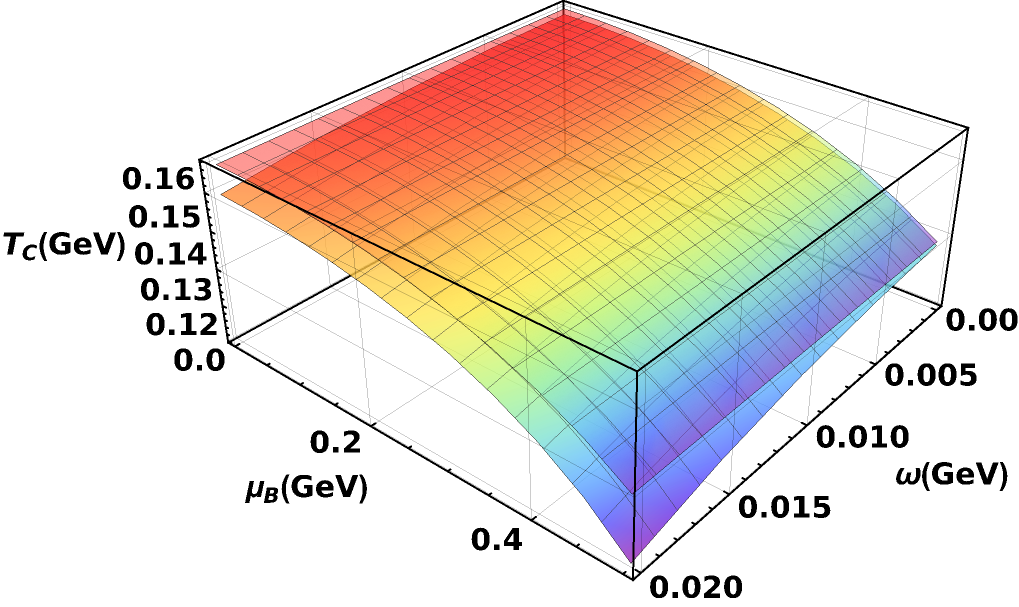}
\includegraphics[width=70mm]{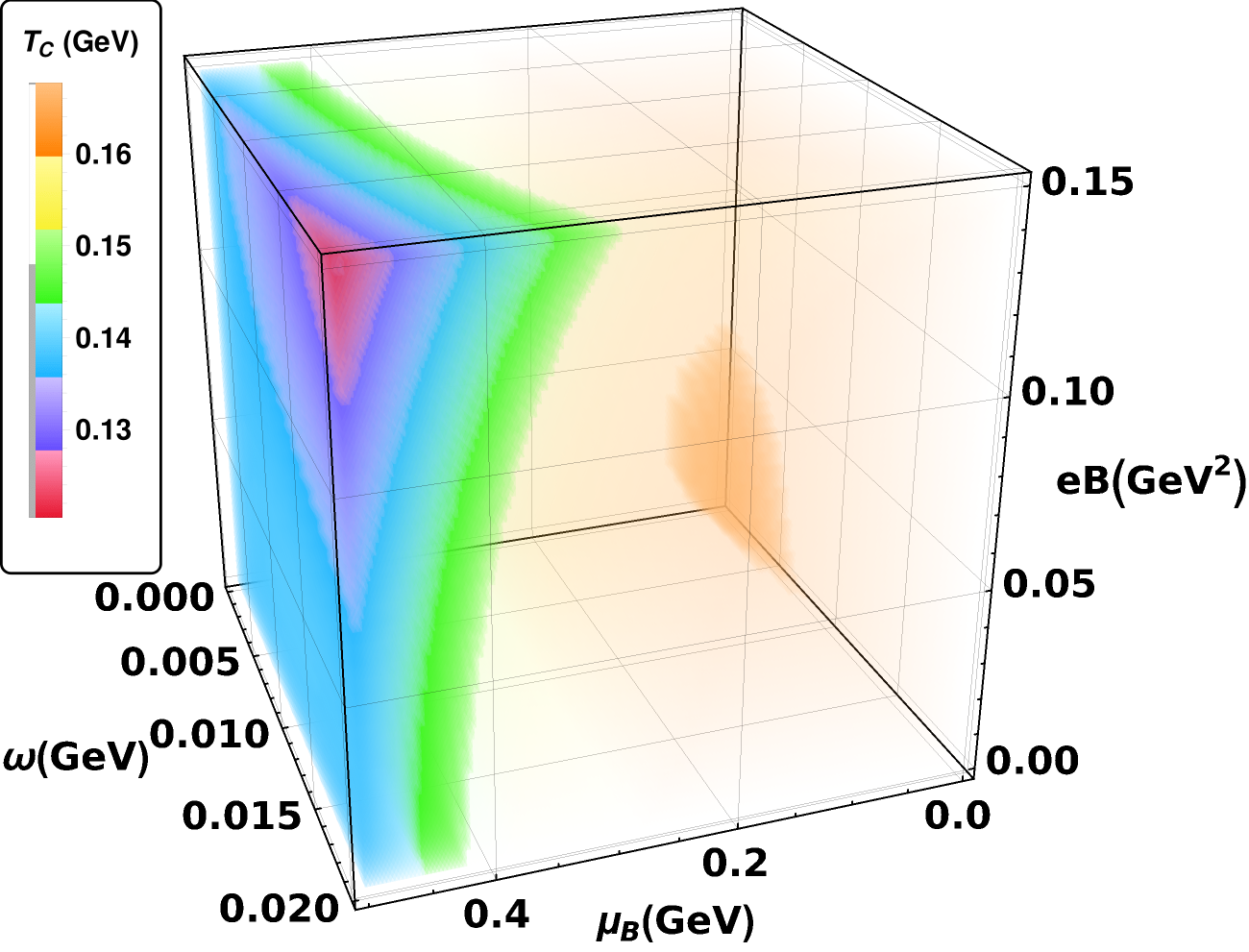}
\caption{\label{fig:3d4d:ent}(Top) Deconfinement transition surfaces showing $T_{C} 
(\mu_B,~\omega)$ for $eB=0$ (upper surface) and $eB=0.15$ GeV$^{2}$  (lower surface). 
(Bottom) Augmented phase diagram showing  $T_{C}(\mu_B,~\omega,~eB)$ as a color-coded 
density plot where the $T_C$-calibrated legend (left) provides reference for the 
different iso-$T_C$ contour boundaries in the $\mu_B,~\omega,~eB$ space. Both plots 
obtained from  rapid rise in entropy density at $s/T^3=5.5$.}
\end{figure}
We observe a successive lowering of the deconfinement crossover zone due to finite 
magnetic field as $\omega$ rises from $0$ to $0.02$ GeV ($\sim 
3\times10^{22}s^{-1}$). The trend is clear across the 
span of these external parameters: the dip in $T_C$ is amplified substantially when 
both the magnetic field and the angular velocity take on high values and at the 
baryon chemical potential of around $0.5$ GeV, the drop reaches a nadir nearing 0.1 
GeV. This is remarkable since at zero magnetic field the effect of rotation with $ 
\omega $ rising from $0$ to $0.02$ GeV is only slight within the range considered here. 
However the simultaneous imposition of an external magnetic field over and above the 
rotation leads to nearly the same drop in the deconfinement temperature at $ \omega $ 
= 0.02 GeV as that estimated at extremely large values of $ \omega $ = 0.3 GeV when 
there is no magnetic field present at all~\cite{Fujimoto:2021xix}. This would suggest 
that although the latter (pure, $eB=0$) rapid rotation scenario might be too high for
 a typical HIC at the time of deconfinement and chemical 
freeze-out, a similar (in magnitude) effective downward shift of the deconfinement 
temperature may nevertheless still apply if the realistic situation sustains a strong 
enough magnetic field ($eB \sim 0.12$ GeV$^2$ 
\cite{Kharzeev:2007jp}) accompanying the more modest but 
also more plausible $\omega$ values ($\sim 0.02$ 
GeV,~\cite{STAR:2017ckg,Becattini:2015ska,Jiang:2016woz,Deng:2016gyh}) as in the case 
of the former which has been examined in the current study.
As the QGP droplet produced in a HIC (particularly those with
finite $\mu_B$, $\omega$ and $eB$) evolves from its initial formation to hadronization,
the pronounced lowering of $T_C$  may lead to a longer
lifetime for the deconfined phase.

In Fig.~\ref{fig:bands}, we have also shown data points for chemical freeze-out
~\cite{STAR:2017prc} as extracted from experimental particle yields and ratiosfor two 
centrality classes, ($0$-$5$)\% and ($60$-$80$)\%. When such fitting analyses 
incorporate rotation and magnetic field as additional quasi-control parameters 
($~\mu_B$, $~\omega$ and$~eB$, all dependent on collision energy and impact 
parameter or centrality), our phenomenological results may be better interpreted. 
The comparison might lead to not only $T-\mu_B$ freeze-out data serving as 
`thermometer' and `baryometer'~\cite{Andronic:2018nat} but also possibly augment them 
with capabilities of `magnetometer'~\cite{Fukushima:2016vix} and `anemometer' to  
estimate the magnitudes of magnetic field and rotational motion prevalent in a HIC 
fireball. The degree of the  relative influences of $\mu_B, \omega$ and $eB$ might be 
constrained from other observable phenomena, for example measured 
polarization~\cite{STAR:2017ckg} to independently constrain $\omega$, etc.
 
Since we now have $T_C(\mu_B,~ \omega,~eB)$, that is the deconfinement temperature 
estimate as a function of three independent parameters, we can depict the variation 
of $T_C$ in higher dimensional spaces to visually understand the individual and 
combined effects of $~\mu_B$, $~\omega$ and$~eB$ on $T_C$. 
Top panel of Fig.~\ref{fig:3d4d:ent} shows the surface plots (3-D QCD phase diagram) 
$T_C(\mu_B,~\omega)$ for $eB=0$ and $eB=0.15$ GeV$^2 $. The values of $T_C$ 
on each surface are obtained from the rapid rise condition of scaled entropy density, 
$s/T^3$ = 5.5. This particular value of the scaled entropy density is found to sit
roughly midway within the deconfinement bands depicted in Fig.~\ref{fig:bands}.
This also leads to a close correspondence in the variation of $T_C$ 
vs. $\mu_B$ as compared with the parallel analysis made using the minima of the 
squared speed of sound. The lowering of the $T_C$ 
vs. $\mu_B$ curve due to the imposed magnetic field of $eB=0.15$ GeV$^2$ is seen 
to be progressively amplified as we crank up 
$\omega$ from $0$ to $0.02$ GeV.
 An even more fine-grained data set is next exhibited
in the (4-D QCD phase diagram) $ T_C (\mu_B, \omega, eB)$ density plot in the bottom
panel of Fig.~\ref{fig:3d4d:ent}.  
The diagram shows a detailed view of the initial gradual $T_C$ change near the origin 
at ($\mu_B,~\omega,~eB)=(0,~ 0,~ 0)$ as the independent variables rise from $0$ 
values and the rapid drop in $T_C$ when the three attain higher values near the 
diagonally opposite corner where isotherms (constant $T_C$ contours) take shape and 
pile up in layered or stratified form. These phase portraits, from the discrete 
curves, bands and surfaces (2-D and 3-D) to the continuous density plots (4-D) help 
us visualize the $T_C$ dependence on increasing $\mu_B,~\omega$ and $eB$ in the 
experimentally interesting parameter space chosen here.

We also reach an independent estimate for the deconfinement temperature by utilizing 
the squared speed of sound which, like the entropy density, is a quantity that is 
insensitive to any vacuum term contribution and can be computed from the modified HRG 
model. A dip in the squared speed of sound, $c_s^{2}$, reveals the softest point in 
the equation-of-state and signals a phase transition. Thus the locations of the 
minima should serve as a faithful proxy for the onset of deconfinement. We exploit 
this property to obtain an independent prediction of the deconfinement region within 
the QCD phase diagram. Figure~\ref{fig:bands} superposes the $T_C$ curves obtained 
from the $c_s^2$ minima onto the results obtained from the Hagedorn limiting 
temperature condition on the entropy density. Both methods lead to the conclusion 
that the drop in $T_C$ is strongly enhanced at high $\mu_B,~\omega$ and $eB$.
\begin{figure}[h]
\includegraphics[width=70mm]{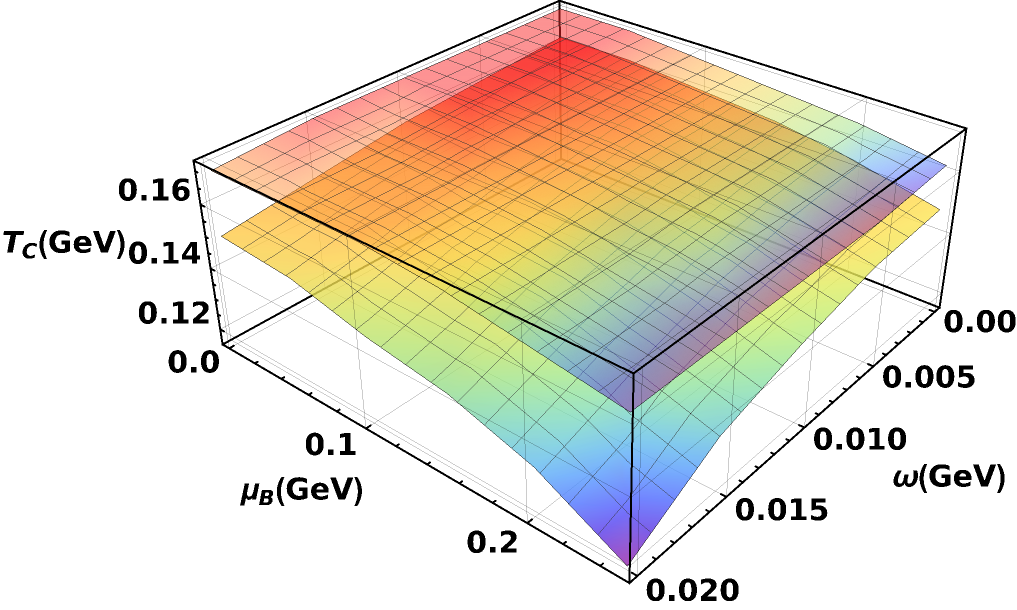}
\includegraphics[width=70mm]{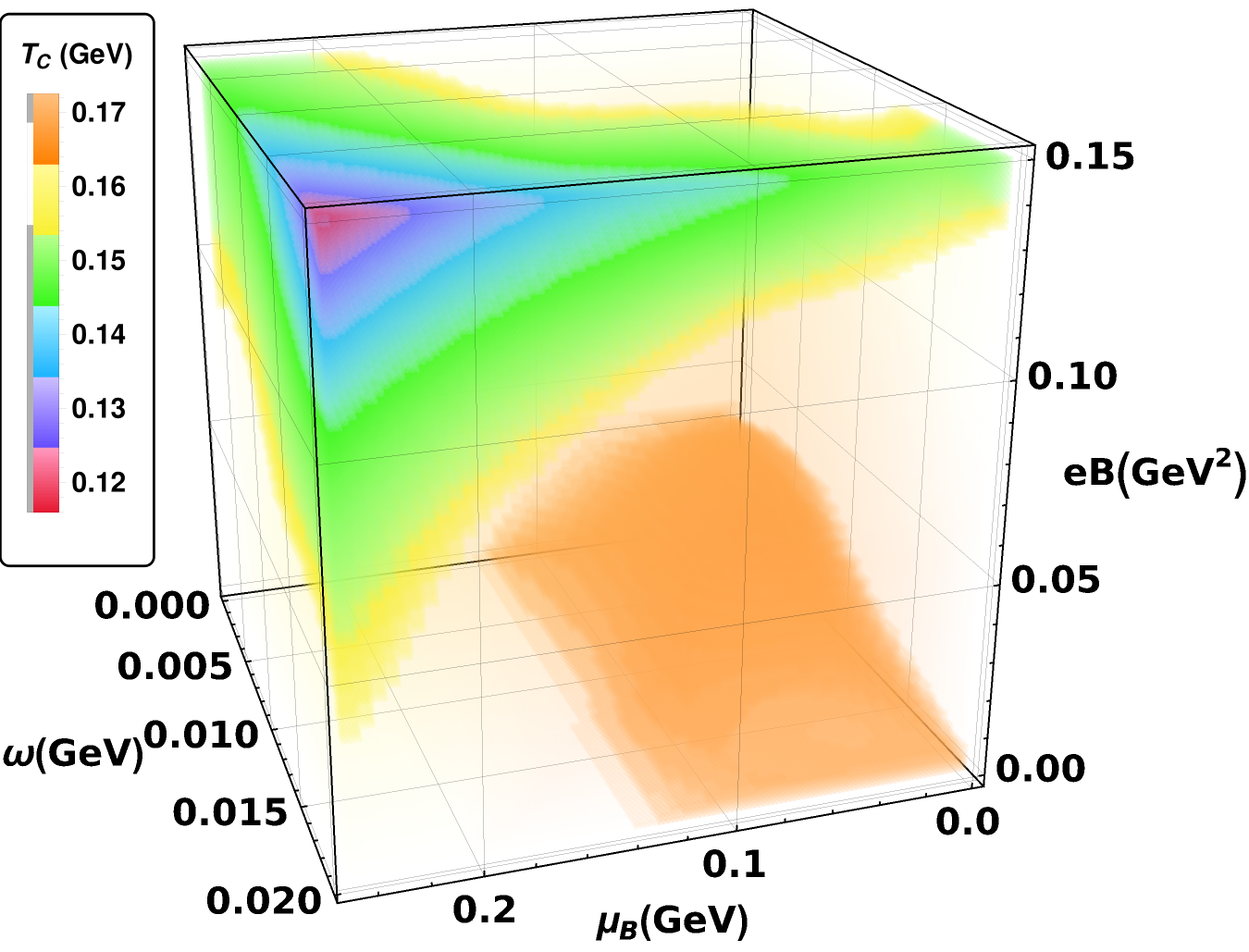}
\caption{\label{fig:3d4d:sos}(Top) Deconfinement transition surfaces showing $ T_{C} 
(\mu_B,~\omega) $ for $eB=0$  (upper surface) and $eB = 0.15$ GeV$^{2}$ 
(lower surface). (Bottom) Augmented phase diagram showing  $T_{C} (\mu_B,~\omega,~eB)$ as a 
color-coded density plot where the $T_C$-calibrated legend (left) provides reference 
for the different iso-$T_C$ contour boundaries in the $\mu_B,~\omega,~eB$ space. Both 
obtained from the minima of the squared speed of sound.}
\end{figure}

Figure~\ref{fig:3d4d:sos} shows the $T_C$ values obtained from the speed of sound 
analysis and these results are analogous to the results shown in 
Fig.~\ref{fig:3d4d:ent}. In our investigations of the squared speed of sound, the 
observed dip becomes shallower~\cite{He:2022kbc,morita} and shows smearing as $\mu_B$ 
or $\omega$ or eB rise. We also notice that the drop in $T_C$ obtained from the 
entropy density analysis is somewhat smaller than that from the speed of sound 
analysis. A similar trend, namely the dip in the squared speed of sound being more 
sensitive (to the magnetic field) than the entropy density, was observed in 
Ref.~\cite{Endrodi:2013cs} also. An exact quantitative correspondence is not expected 
anyway between the results from the two distinct methods and thus the obtained 
pseudo-critical temperatures (no unique $T_C$ for a non-singular crossover 
\cite{Aoki:2006br}) serve as an approximate determination of a continuous transition. 
The ambiguity of precisely where the deconfinement curve/surface cuts the $T$ axis 
notwithstanding, we infer that the augmented HRG model predicts a decrease in $T_C$ 
(by a maximum of 40 to 50 MeV in the parameter ranges considered here) as we venture 
into the (non-zero $T$,~$\mu_B,~\omega,~eB$) hinterland of the augmented QCD phase 
space.

We now discuss the physics motivations behind taking two distinct routes. Firstly, 
since there is no treatment in the literature, to the best of our knowledge, that 
studies the non-perturbative QCD thermodynamics and phase structure under 
simultaneously non-zero $\omega$, $\mu_B$ and $eB$, it is important to cross-check 
the results we obtain by confirming that our two different approaches are mutually 
validated. If the results obtained by constraining $s/T^3$ (invoking the Hagedorn 
argument) are corroborated qualitatively and even quantitatively to close proximity 
with an independent method, like minimizing $c_s^2$, one can be confident that the 
consistent results from these methods constitute a robust and novel prediction of the 
augmented HRG model. Thus our twin approach serves to substantiate the basic results 
found. Again, the choice of $s/T^3$ and $c_s^2$ over other criteria like scaled 
$p$~\cite{Fujimoto:2021xix} and $n$~\cite{Fukushima:2010is}, $\epsilon / 
n$~\cite{Fukushima:2016vix,PhysRevC.99.024902}, 
$n_B+n_{\overline{B}}$~\cite{PhysRevC.99.024902} lets us bypass the vacuum term 
($T=0$) that would otherwise require renormalization~\cite{Endrodi:2013cs}. Also, 
$c_s^2$ as computed in the generalized HRG model treated here opens up the 
possibility of going beyond studying just the freeze-out stage thermodynamics to also 
matching with the hydrodynamics of the QGP phase. Our evaluation of $c_s^2$, being a 
key ingredient in hydrodynamic calculations, fireball evolution, elliptic flow 
studies, conformal symmetry breaking in hot QCD, etc. \cite{Khaidukov:2018lor} will 
serve as a starting point for more detailed studies in forthcoming work. 

We discuss the HIC phenomenological aspects with regard to the `magnetometer' and 
`anemometer' measurability proposal. Let us denote by `x' the control parameters 
(like collision energy, centrality, rapidity) that can be adjusted to vary the 
$\mu_B$, $\omega$ and $eB$ at which the system crosses the transition 
region~\cite{Stephanov:1998dy}. The existing parameterization for chemical freeze-out 
data seems well-suited to be extended to include $\omega$, $eB$ along with $T$ and 
$\mu_B$. The universal freeze-out condition, $s/T^3 \sim 7$~\cite{CLEYMANS200550}, 
can be tested thereafter in the extrapolated phase space by comparing with $T-\mu_B$ 
data for varying centrality, collision energy and rapidity. The $c_s^2$ minima are 
more directly associated with QCD phase transitions, particularly the deconfinement 
boundary~\cite{Khaidukov:2018lor}. There is strong evidence that the deconfinement 
boundary and chemical freeze-out line merge for small $\mu_B$. This is supported by 
the common and close trend found in our results in Fig.~\ref{fig:3d4d:ent} and 
Fig.~\ref{fig:3d4d:sos}. The quantitative differences between Fig.~\ref{fig:3d4d:ent} 
and Fig.~\ref{fig:3d4d:sos} will probably help in the quite non-trivial task of 
disentangling or segregating the individual contributions of $\mu_B$, $\omega$ and 
$eB$ to the convoluted processes within a hadronizing HIC droplet. Perhaps some 
additional refinements in modeling the actual fireball system more realistically 
will provide further fine-tuning of the results in the future.  

It is interesting to note that certain aspects of the chiral transition explored in 
the literature~\cite{Wang:2018sur,Chernodub:2017ref} share similarities to the 
deconfinement transition which has been the focus of our study here. The inverse 
magnetic catalysis (IMC) leads to a decrease in the chiral transition temperature 
with an increasing magnetic field in the presence of non-zero chemical potential and 
also some other influences~\cite{Fukushima:2012kc,Bali:2011qj,
Preis:2012fh,Bruckmann:2013oba}. Observing that rotation induces an effective 
chemical potential, a phenomenon analogous to inverse magnetic catalysis called 
rotational magnetic inhibition~\cite{Chen:2015hfc} or inverse magnetorotational 
catalysis (IMRC)~\cite{Sadooghi:2021upd} was proposed. It is conceivable that the 
common decreasing trend expected for the critical temperature in these studies 
involving, in each case, only two of the three independent variables 
$\mu_B,~\omega,~eB$, gets reinforcement when all of the three are present 
simultaneously. This work may shed some light on whether the two transition 
temperatures stay locked in value or split as we turn on the various parameters to 
finite values and advance into the (augmented) phase diagram interior.

\section{Summary}
\label{sec:sum}
We set out to examine the deconfinement zone of the QCD phase diagram for hot and 
dense matter that is subjected to an external magnetic field and a parallel global 
rotation as might be present in a typical non-central heavy-ion collision. For this 
we utilized the HRG model with suitable modifications due to the additional 
parameters of magnetic field and angular velocity. We calculated the entropy density 
and squared velocity of sound as functions of temperature. We showed, based on our 
imposed criteria for the identification of deconfinement, that the simultaneous 
turning on of both the magnetic field and angular velocity appears to significantly 
amplify the drop in $ T_C $ due to baryon-chemical potential even further, by close 
to 40 to 50 MeV. We observed that the decrease of $T_C$  for HIC values of 
$\mu_B,~\omega$ and $eB$ is substantial and comparable to that due to the much 
higher but experimentally implausible $eB$ or $\omega$ values if considered 
separately. One potential application is the possibility of accessing the HIC 
fireball properties by using this approach as a `thermometer', `baryometer', 
`magnetometer' and `anemometer'-like tool. We also discussed a possible extension of 
the lifetime of the QGP phase in HIC, the deconfinement and chiral transitions in our 
new regime and speculated about their correspondence, i.e., splitting versus locking. 
The direction of research pursued here, incorporating the combined effects of 
$\mu_B,~\omega$ and $~eB$, opens up new  avenues for future explorations into 
uncharted QCD phase territory, in tandem with present and upcoming heavy-ion 
colliders.      
           
\noindent {\bf Acknowledgments }
 The authors thank A. K. Mohanty for valuable feedback and discussions. We also 
 gratefully acknowledge the anonymous referee whose insightful suggestions led to 
overall improvement of this work.

\bibliographystyle{elsarticle-num}
\bibliography{phase_diagram}

\end{document}